\documentclass[
twocolumn,
prl,
amsmath,
amssymb,
superscriptaddress
]{revtex4-2}

\usepackage{graphicx}% Include figure files
\usepackage{dcolumn}% Align table columns on decimal point
\usepackage{bm}% bold math
\usepackage{subfigure}
\usepackage{hyperref}
\usepackage{amsmath}
\usepackage{hyphenat}
\usepackage{amsfonts}
\usepackage{booktabs}
\usepackage{multirow}
\usepackage{textcomp}
\usepackage[usenames]{color}
\hypersetup{bookmarks=true, colorlinks=true,linktoc=page, linkcolor=blue, citecolor=blue, urlcolor=blue }
\graphicspath{%
}%

\begin{document}

\preprint{APS/123-QED}

\title{Constraint of the Nuclear Dissipation Coefficient in Fission of Hypernuclei
}% Force line breaks with \\
\author{J. L. Rodr\'{i}guez-S\'{a}nchez}
\email[Corresponding author: ]{j.l.rodriguez.sanchez@udc.es}
\affiliation{CITENI, Campus Industrial de Ferrol, Universidade da Coru\~{n}a, E-15403 Ferrol, Spain}
\affiliation{IGFAE, Universidad de Santiago de Compostela, E-15782 Santiago de Compostela, Spain}
\author{J. Cugnon}
\affiliation{AGO department, University of Li\`{e}ge, all\'{e}e du 6 ao\^{u}t 19, b\^{a}t.~B5, B-4000 Li\`{e}ge, Belgium}
\author{J.-C. David}
\affiliation{IRFU, CEA, Universit\'{e} Paris-Saclay, F-91191 Gif-sur-Yvette, France}
\author{J. Hirtz}
\affiliation{IRFU, CEA, Universit\'{e} Paris-Saclay, F-91191 Gif-sur-Yvette, France}
\affiliation{Physics Institute, University of Bern, Sidlerstrasse 5, 3012 Bern, Switzerland}
\author{A. Keli\'{c}-Heil}
\affiliation{GSI-Helmholtzzentrum f\"{u}r Schwerionenforschung GmbH, D-64291 Darmstadt, Germany}
\author{I.~Vida\~{n}a}
\affiliation{INFN, Sezione di Catania, Dipartimento di Fisica ``Ettore Majorana", Universit\`a di Catania, I-95123 Catania, Italy}

\date{\today}% It is always \today, today,

\begin{abstract}
Experimental studies of nuclear fission induced by fusion, transfer, spallation, fragmentation, and electromagnetic reactions in combination with state-of-the-art calculations are successful to investigate the nuclear dissipation mechanism in normal nuclear matter, containing only nucleons. The dissipation mechanism has been widely studied by the use of many different fission observables and nowadays the dissipation coefficients involved in transport theories are well constrained. However, the existence of hypernuclei and the possible presence of hyperons in neutron stars make it necessary to extend the investigation of the nuclear dissipation coefficient to the strangeness sector. In this Letter, we use fission reactions of hypernuclei to constrain for the first time the dissipation coefficient in hypernuclear matter, observing that this coefficient increases a factor of 6 in presence of a single $\Lambda$-hyperon with respect to normal nuclear matter.
\end{abstract}
\maketitle
\textit{Introduction.--} 
%{\color{blue}
Dissipation mechanisms play an important role in nuclear physics and astrophysics to describe the dynamics of nuclei and neutron stars (NSs), respectively. In nuclear physics one of the most intensively investigated nuclear dynamics mechanism involving dissipation is the fission process, in which a heavy nuclear system is deformed until it splits into two lighter  fragments with similar masses~\cite{Meitner1939,Hahn1939}. The general picture of this decay process naturally leads to the fission description in terms of a nuclear potential-energy surface as a function of the nuclear shape~\cite{Moller2001}, but its complete modeling also requires the knowledge of dynamic properties of the fissioning system, namely, the static nuclear configurations out of equilibrium, the coupling between collective and intrinsic degrees of freedom, and the dynamics of large amplitude collective motion. The dynamical evolution through the nuclear potential-energy landscape is dominated by the exchange of energy between the collective and intrinsic degrees of freedom, in which the transfer of energy is described as a dissipative process. According to a semiclassical picture, the most complete transport treatment is provided by approaches of motion based on the Langevin~\cite{Wada1992,Sierk2017} or Fokker-Planck equation (FPE)~\cite{Risken1989}, where the viscosity coefficient appears as a free parameter. The collective coordinates usually exhibit a Brownian-like motion that can be simulated numerically as a random walk~\cite{Grange1980,Randrup2011}. The simplicity of this approach together with its remarkable agreement with known data make it suitable for global studies of fission probabilities and other fission observables~\cite{Frobrich1988,Nadtochy2007,Randrup2013}.

The presence of hyperons (baryons with strange content) in finite and infinite nuclear systems, such as hypernuclei and neutron stars, make it necessary to extend the investigation of the nuclear dissipation coefficient to the strangeness sector. Hypernuclei, bound systems composed of nucleons and hyperons, can be produced by several reactions (see e.g. Ref.~\cite{Gal2016}) where hyperons can be captured by nuclei since their lifetimes ($\sim$ hundreds of ps) are longer than the characteristic reaction times ($\sim10^{-23}$~s)~\cite{Park2000}. 
The study of hypernuclei and their properties provides the opportunity to study baryon-baryon interactions \cite{Millener1988,Maessen1989,Tominaga98,Polinder2006,Haidenbauer2013,Hiyama2018,Haidenbauer2020} from an enlarge perspective and to extend, in this way, our present knowledge of conventional nuclear physics to the SU(3)-flavor sector~\cite{Lenske2018}.

In astrophysics the dissipation mechanisms also play a crucial role in the understanding of the  oscillation modes of NSs, which have attracted interest for a considerable time. Significant effort has been aimed at understanding whether gravitational-wave emission sets the upper rotational frequency limit for pulsars, e.g., via the r-mode instability discovered by Andersson, Friedman, and Morsink~\cite{Andersson1998,Friedman1998} in 1998. This possibility is of particular interest since r-modes can lead to the emission of detectable gravitational waves in hot and rapidly rotating NSs. R-modes are predominantly toroidal oscillations (i.e., oscillations with a divergenceless velocity field and a suppressed radial component of the velocity) of rotating stars restored by the Coriolis force~\cite{Kolomeitsev2015}, acting similar to Rossby waves in Earth’s atmosphere and oceans. It is of great importance to understand whether internal fluid dissipation allows the instability to develop in such systems or whether it suppresses the r-modes completely. There are many mechanisms in real NSs that compete with the gravitational wave driving of the r-mode. The instability can only develop when the gravitational radiation growth timescale is shorter than the damping timescales due to the various viscosity mechanisms, such as exotic bulk viscosity due to the presence of hyperons or quark matter~\cite{Madsen1998,Lindblom2002}, and enhanced mutual-friction dissipation~\cite{Lindblom2000}. This defines a region in the spin-temperature parameter space where the r-mode instability is active. If the instability occurs, the star rapidly radiates its angular momentum via gravitational waves until the rotation frequency reaches the critical value, at which the r-modes become stable. This rapid process is accompanied by the NS reheating~\cite{Alford2012}. Therefore, the dissipation process suppresses this instability and makes cold and hot NSs effectively stabilized by shear and bulk viscosities~\cite{Andersson2001,Lindblom2002,Nayyar2006}, respectively. Unfortunately, direct measurements of the shear and bulk viscosities in NSs do not exist yet.

In this Letter, we propose to investigate the dissipation coefficient in hypernuclear matter following the methodology proposed in pioneering works on nuclear fission, in which fission cross sections~\cite{JL2014,Ayyad2015}, charge distribution of fission fragments~\cite{Ayyad2015,JL2015,CS2007}, multiplicity of $\gamma$-rays, neutrons, and light-charged particles~\cite{Frobrich1988,JL2016} were used to obtain the nuclear dissipation coefficient in normal nuclear matter. For the present work we take advantage of the data collected during the hypernuclei experiments~\cite{Ohm1997,Kulessa1998} performed at the COSY-J\"{u}lich facility (Germany) by comparing the fission cross sections induced in heavy hypernuclei to dynamical calculations based on the FPE approach, which also allows us for the first time to study the dependence of the dissipation parameter on the number of hyperons that constitute the nuclear system.

\textit{Theoretical framework.--}
The collision between the proton and target nuclei, the so-called spallation reaction, is described with the latest C++ version of the dynamical Li\`{e}ge intranuclear-cascade model (INCL)~\cite{Mancusi14} coupled to the ablation model ABLA~\cite{JL2022}, which are based on Monte Carlo techniques obeying all conservation laws throughout each reaction event. INCL describes the spallation reaction as a sequence of binary collisions between the nucleons (hadrons) present in the system. Nucleons move along straight trajectories until they undergo a collision with another nucleon or until they reach the surface, where they could possibly escape. The latest version of the INCL also includes isospin- and energy-dependent nucleus potentials calculated according to optical models~\cite{Bou13}, isospin-dependent pion potentials~\cite{Aoust2006}. INCL has been recently extended toward high energies ($\sim$20 GeV) including new interaction processes, such as multipion production~\cite{Mancusi2017}, production of $\eta$ and $\omega$ mesons~\cite{JCD2018}, and strange particles~\cite{Jason2018,JL2018,Jason2020}, such as kaons and hyperons. Therefore this new version of INCL allows us to predict the formation of hot hyperremnants and their characterization in atomic ($Z$) and mass ($A$) numbers, strangeness number, excitation energy, and angular momentum. We remark that the good agreement of INCL calculations with experimental kaon production cross sections obtained from proton-induced reactions on light, medium-mass, and heavy nuclei at energies of few GeV~\cite{Jason2020}, as well as the reasonable description of hypernuclei production cross sections through strangeness-exchange reactions ($\pi^{+}$,$K^{+}$)~\cite{JL2018}, allow us to guarantee a correct prediction of the excitation energy gained by the hyperremnants after the proton-nucleus collision.

The hyperremnants enter then the deexcitation stage that is modeled by the code ABLA. This model describes the deexcitation of a nuclear system through the emission of $\gamma$-rays, neutrons, $\Lambda$-hyperons, light-charged particles, and intermediate-mass fragments (IMFs) or fission decays in case of hot and heavy compound nuclei. The particle emission probabilities are calculated according to the Wei$\beta$kopf-Ewing formalism~\cite{We40}, being the separation energies and the emission barriers for charged particles calculated according to the atomic mass evaluation from 2016~\cite{mass2016} and the prescription given by Qu and collaborators~\cite{Qu2011}, respectively. The fission decay width is described by the Bohr-Wheeler transition-state model~\cite{BW39}, following the formulation given by Moretto and collaborators~\cite{Moretto75}:
\begin{equation}\label{eqbw}
\Gamma_{f}^{BW}= \frac{1}{2 \pi \rho_{gs}(E,J)}
\int_{0}^{E-B_f} \rho_{sp}(E-B_f-\epsilon,J) d\epsilon,
\end{equation}
where $\rho_{gs}$ and $\rho_{sp}$ are the level densities at the ground state and the fission saddle point, respectively, E is the excitation energy, J represents the angular momentum, and B$_f$ is the fission-barrier height obtained from the finite-range liquid-drop model of Sierk~\cite{Sierk86} taking into account the influence of angular momentum and considering the ground-state shell effects~\cite{Moller95}. In the case of hypercompound systems, the hyperenergy contribution $\Delta B_f^{hyp}$ released during the fission process due to the attractive YN force is added to $B_f$ in Eq.~\ref{eqbw}. This hyperenergy $\Delta B_f^{hyp}$ is parametrized according to the description given by Ion and collaborators~\cite{Ion1989} as follows:
\begin{equation}
\Delta B_f^{hyp} = 0.51( m_{\Lambda} - m_{n} + S_{n} - S_{\Lambda})/A^{2/3},\nonumber
\end{equation}
where m$_{\Lambda}$ (S$_{\Lambda}$) and m$_{n}$ (S$_{n}$) are the $\Lambda$ and neutron masses (separation energies), respectively, and $A$ is the mass number of the hypercompound nucleus. This equation leads to small increases in the fission barrier height up to $\Delta B_f \sim$ 1 MeV that are, for instance, compatible with the results obtained from more sophisticated calculations based on Skyrme-Hartree-Fock approaches~\cite{Minato2009}.

The slowing effects of nuclear dissipation are included by using the
Kramers-modified Bohr-Wheeler model~\cite{KR40} as follows:
\begin{equation}\label{eqkramers}
\Gamma_{f}^{K} = \left[ \sqrt{1+ \left(\frac{\beta}{2 \omega_{0}}\right)^2} -\frac{\beta}{2 \omega_{0}} \right] \Gamma_{f}^{BW},
\end{equation}
where $\beta$ is the nuclear dissipation coefficient and $\omega_{0}$ is the frequency of the harmonic oscillator describing the inverted potential at the fission barrier, calculated according to the liquid-drop model~\cite{Nix1967}. This equation provides the asymptotic value of the fission decay width, which is modified according to an analytical approximation to the solution of the one-dimensional Fokker-Planck equation~\cite{Risken1989}, developed by Jurado and collaborators in Refs.~\cite{BJ03,BJ05}, to take into account the time dependence of the fission-decay width. The Eq.~\ref{eqkramers} was modified later to consider the initial quadrupole deformation of a compound nucleus, which provides a more realistic description of fission yields in the actinide region~\cite{CS2007}. Under this approximation, the time-dependent fission-decay width is given as:
\begin{equation}
\Gamma_f(t)= \frac{W_n (x = x_{sd} ; t, \beta)}{W_n (x = x_{sd} ; t\rightarrow \infty, \beta)} \Gamma_{f}^{K},\nonumber
\end{equation}
where $W$($x; t, \beta$) is the normalized probability distribution at the saddle-point deformation $x_{sd}$, being the saddle-point deformations calculated according to Ref.~\cite{Hasse}.

\textit{Application to proton-induced fission reactions on hypernuclei.--}
In the 1990's an experimental campaign was carried out at the COSY-J\"{u}lich facility to measure the lifetime of $\Lambda$-hyperons in the nuclear medium using fission reactions of heavy hypernuclei as a tool to reconstruct the $\Lambda$-hyperon decay vertex. The measurements were performed by using the recoil shadow method~\cite{Ohm1997,Kulessa1998}, which also allowed for the first time to determine the hypernuclear fission cross sections for two nuclear systems: $^{238}$U and $^{209}$Bi. 

\begin{table}[b!]
\caption{Results obtained for the dissipation parameter on normal and hypernuclear matter. 
}
\label{tab1}
\begin{center}
\begin{ruledtabular}
\begin{tabular}{cc}
Reaction &  $\beta$ [$10^{21}$ s$^{-1}$] \\
\noalign{\smallskip}\hline\noalign{\smallskip}
Spallation and fragmentation of normal nuclei &  4.5 $\pm$ 2.0\\
p($^{209}$Bi,hypernuclei) & 16 $\pm$ 7\\
p($^{238}$U, hypernuclei)  & 40 $\pm$ 10\\
\end{tabular}
\end{ruledtabular}
\end{center}
\end{table}

\begin{figure}[t!]
\centering
\subfigure{\label{fig:1a}\includegraphics[width=0.49\textwidth,keepaspectratio]{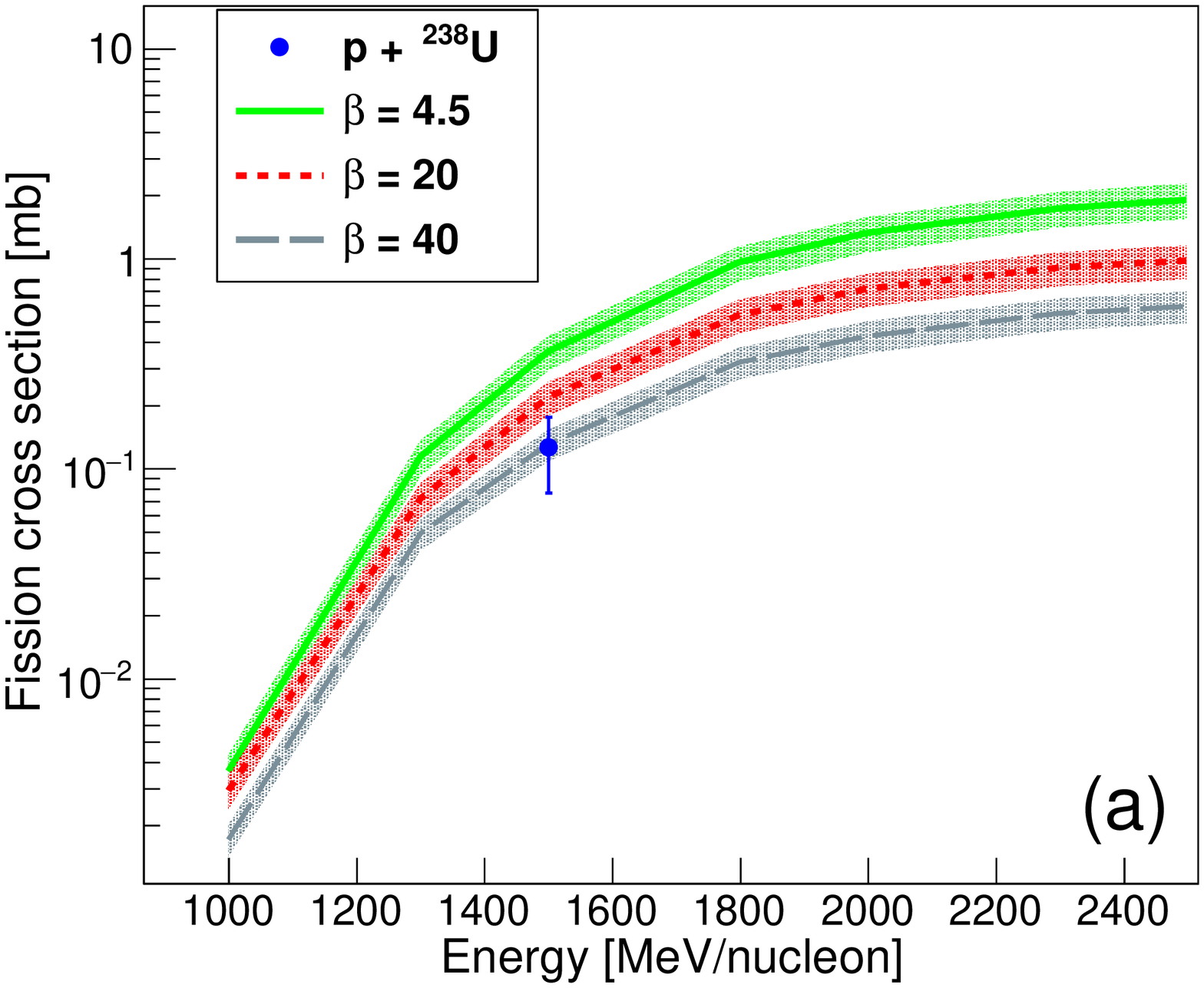}}
\centering
\subfigure{\label{fig:1b}\includegraphics[width=0.49\textwidth,keepaspectratio]{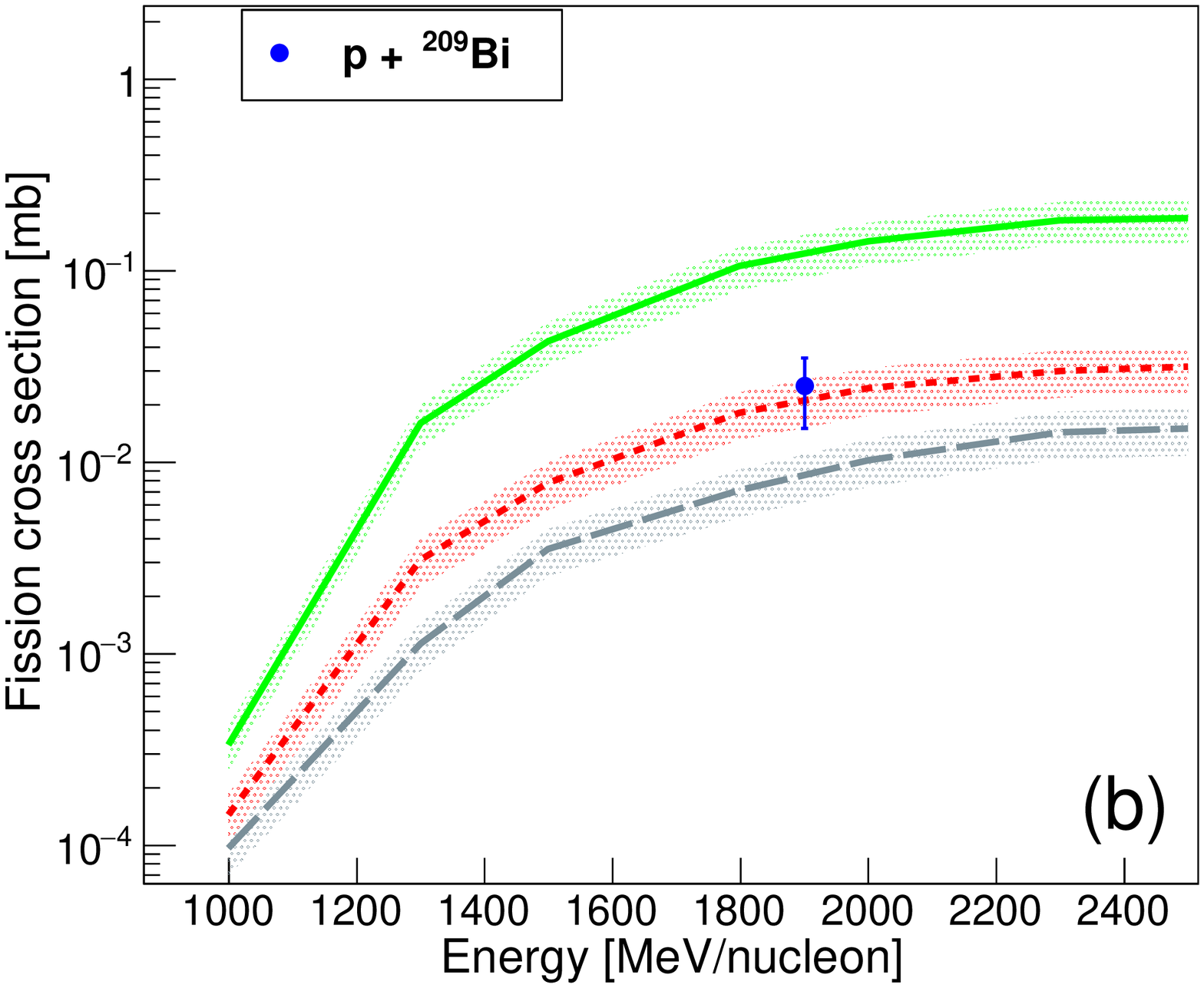}}
\caption{(Color online) Hypernuclear fission cross section (dots) as a function of the projectile kinetic energy per nucleon for target nuclei of $^{238}$U (a) and $^{209}$Bi (b). The lines correspond to dynamical fission calculations for different values of the dissipation coefficient $\beta$. Dashed areas represent the uncertainties (see text for details).
}
\label{fig:1}
\end{figure}

In Figs.~\ref{fig:1a} and \ref{fig:1b} we display the experimental cross sections obtained for hypernuclear fission reactions induced in $^{238}$U and $^{209}$Bi nuclei, respectively, as a function of the proton kinetic energy. These data are compared to INCL+ABLA calculations in which we have assumed different values for the dissipation parameter $\beta$: 4.5 (solid line), 20 (short-dashed line), and 40 (long-dashed line)$ \times 10^{21}$ s$^{-1}$. We can see that the hypernuclear fission cross sections decrease when increasing the value of the viscosity parameter, which is expected since the nuclear system evolves more slowly, needing more time to reach the saddle point configuration. This fact favors the cooling of the nuclear system by particle emission reducing the fission probabilities. In these calculations we also take into account the uncertainties in the nuclear level densities and fission barrier heights, which are displayed in the figures with dashed areas. These uncertainties do not exceed 18$\%$ of the total hypernuclear fission cross section, being the 16$\%$ of this uncertainty attributed to the fission barrier height. The comparison allows us to constrain the value of the viscosity coefficient for both nuclear systems, resulting in the values given in Table~\ref{tab1}, where the uncertainties were calculated by propagating the uncertainty of each experimental fission cross section. The values are also compared to that obtained from spallation and fragmentation reactions of normal nuclei inducing fission.

\begin{figure}[h!]
\begin{center}
\includegraphics[width=0.49\textwidth,keepaspectratio]{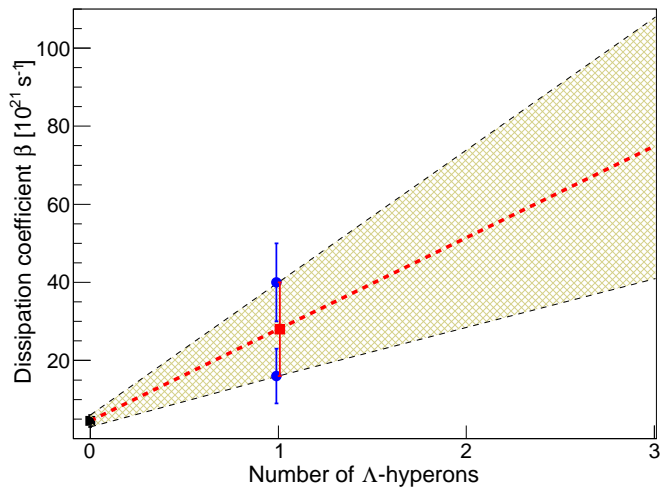}
\caption{(Color online) Dissipation coefficient as a function of the number of $\Lambda$-hyperons presented in the nuclear system. The values for a single $\Lambda$-hyperon (dots) are extracted from the constraints displayed in Fig.~\ref{fig:1}. We also display the average dissipation coefficient for hypernuclear matter with a $\Lambda$-hyperon (square). The dashed area indicates the expected region for the dissipation coefficient at larger $\Lambda$-hyperon multiplicities, where the red dashed line indicates the linear extrapolation of the average values obtained for normal nuclear matter and hypernuclear matter with a single $\Lambda$-hyperon.
}
\label{fig:2}
\end{center}
\end{figure}

The results of Table~\ref{tab1} are also displayed in Fig.~\ref{fig:2} to illustrate the evolution with the presence of $\Lambda$-hyperons in the nuclear system. Here we believe the nuclear dissipation coefficient is well constrained for normal nuclear matter taking into account the large number of nuclear fission experiments performed in the last decades, so the value of $(4.5 \pm 2.0) \times 10^{21}$ s$^{-1}$ is quite robust~\cite{JL2014,Ayyad2015,Ayyad2015,JL2015,CS2007,JL2016,JB2017,Rodriguez2016}. Applying the same calculations to hypernuclear matter we find that the value of the dissipation coefficient increases roughly a factor of 6 due to the presence of a $\Lambda$-hyperon and the linear extrapolation of our findings to a larger number of $\Lambda$-hyperons leads to higher dissipation coefficients, for instance, $\beta = 75 \times 10^{21}$ s$^{-1}$ for hypernuclei containing three $\Lambda$-hyperons.

We therefore conclude that the dissipation coefficient increases with the presence of hyperons, but we would like to remark here that for the nuclei used in this work there is only a single experimental data point and, thus, more measurements have to be performed to complete the evolution of the hypernuclei fission cross sections with the proton kinetic energy.

\textit{Summary and conclusions.--} 
The hypernuclei fission cross sections measured at the COSY-J\"{u}lich facility with high energetic protons impinging onto target nuclei of $^{209}$Bi and $^{238}$U are used for the first time to investigate the nuclear dissipation coefficient in the presence of hypernuclear matter. The experimental data were compared to a state-of-the art dynamical model that describes the proton-nucleus collision and the subsequent deexcitations through particle and cluster emission as well as nuclear fission. The comparison allowed us to constrain the dissipation parameter in fission of hypernuclear matter, resulting in an average value of $(28 \pm 12) \times 10^{21}$~s$^{-1}$~\cite{std}. This finding is 6  times larger than that obtained for normal nuclear matter~\cite{Nadtochy2007,JL2014,Ayyad2015,JL2016,JB2017,Rodriguez2016}, which implies that in presence of hyperons the conversion of intrinsic energy into collective motion goes much slower.

We also conclude with this work that more experimental data are required to confirm the present results. This lack may be overcome with the construction of new facilities, such as the Facility for Antiproton and Ion Research (FAIR)~\cite{Fairweb} at Darmstadt (Germany) and the High Intensity Heavy-ion Accelerator Facility (HIAF)~\cite{Hiafweb} at Huizhou (China), together with the design of state-of-the art detectors for hypernuclear matter research and the increase of luminosity for the production of relativistic radioactive beams. The forthcoming experiments could provide unique opportunities to study hypernuclear matter in a large range of atomic numbers, from light to heavy nuclei, and to investigate hypernuclei in extreme proton-neutron asymmetry conditions far away from the stability valley. 

Measurement of other fission observables, such as the charge and mass distributions of the fission fragments as well as the multiplicity of emitted particles, in combination with relativistic proton-nucleus and nucleus-nucleus~\cite{hades} collisions in inverse kinematics might shed light on the dependence of the dissipation parameter with the nuclear density and nuclear deformations, as already investigated with normal nuclei~\cite{CS2007,JL2015,JL2016,Rodriguez2016}. Moreover, the measurement of double-$\Lambda$ hypernuclei fission reactions could be used to constrain the dissipation coefficient at a larger number of $\Lambda$-hyperons and to better understand the tendency shown in the Fig.~\ref{fig:2}. Therefore these kinds of experimental measurements could be helpful to extrapolate the values of the dissipation coefficient to hyperon matter in neutron stars.

\textit{Acknowledgments.--}
We thank Dr. Alain Boudard, Dr. Davide Mancusi, and Dr. Sylvie Leray for enlightening discussions and Dr. Georg Schnabel for his technical support. This work was partially supported by the P2IO LabEx (ANR-10-LABX-0038) in the framework "Investissements d'Avenir" (ANR-11-IDEX-0003-01), managed by the Agence Nationale de la Recherche (ANR) (France), and by the EU ENSAR2 FP7 project (Grant Agreement No. 654002). I.V. thanks the support of the European Union’s Horizon 2020 research and innovation programme under Grant Agreement No. 824093. J.L.R.-S. is thankful for the support by the Department of Education, Culture and University Organization of the Regional Government of Galicia under the Postdoctoral Fellowship Grant No. ED481D-2021-018 and by the "Ram\'{o}n y Cajal" programme under the Grant No. RYC2021-031989-I, funded by MCIN/AEI/10.13039/501100011033 and by “European Union NextGenerationEU/PRTR”.



\begin{thebibliography}{20}
\bibitem{std}
The uncertainty is calculated with the standard deviation of the mean, using the results listed in the Tab.~\ref{tab1} for hypernuclear matter.
\end{thebibliography}
\end{document}